\documentclass{pasa}%

\usepackage{graphicx}

\usepackage{url}

\usepackage{aas_macros}
\usepackage{hyperref}

\title[Galaxy spin pattern asymmetry]{Large-scale photometric asymmetry in galaxy spin patterns}

\author[Lior Shamir]{Lior Shamir
\affil{Lawrence Technological University, Southfield, MI, 48075}
}

\jid{PASA}
\doi{10.1017/pas.\the\year.xxx}
\jyear{\the\year}

\hypersetup{colorlinks,citecolor=blue,linkcolor=blue,urlcolor=blue}

\hypersetup{draft}

\begin{document}

\begin{frontmatter}
\maketitle

\begin{abstract}
Spin patterns of spiral galaxies can be broadly separated into galaxies with clockwise (Z-wise) patterns and galaxies with counterclockwise (S-wise) spin patterns. While the differences between these patterns are visually noticeable, they are a matter of the perspective of the observer, and therefore in a sufficiently large universe no other differences are expected between galaxies with Z-wise and S-wise patterns. Here large datasets of spiral galaxies separated by their spin patterns are used to show that spiral galaxies with Z-wise spin patterns are photometrically different from spiral galaxies with S-wise patterns. That asymmetry changes based on the direction of observation, such that the observed asymmetry in one hemisphere is aligned with the inverse observed asymmetry in the opposite hemisphere. The results are consistent across different sky surveys (SDSS and PanSTARRS) and analysis methods. The proximity of the most probable asymmetry axis to the galactic pole suggests that the asymmetry might be driven by relativistic beaming. Annotated data from SDSS and PanSTARRS are publicly available.
\end{abstract}

\begin{keywords}
Galaxies: general -- galaxies: photometry -- galaxies: spiral
\end{keywords}
\end{frontmatter}


\section{INTRODUCTION}
\label{introduction}

The position of a face-on spiral galaxy compared to the position of an observer is reflected by the spin patterns of the galaxy, which can seem clockwise (Z) or counterclockwise (S) based on the location of the observer in the universe. Since the spin pattern is merely a matter of the position of the observer, in a sufficiently large population of spiral galaxies, galaxies with Z-wise spin patterns are expected to be identical in their physical properties to galaxies with S-wise spin patterns. However, previous experiments using data from the Sloan Digital Sky Survey \citep{york2000sloan} show that spiral galaxies with Z-wise spin patterns in SDSS have photometric information that makes them distinguishable from galaxies with S-wise spin patterns \citep{sha16,shamir2016photometric}. 

First results \citep{sha16} using 13,440 galaxies from Galaxy Zoo 2 \citep{willett2013galaxy} separated manually to Z-wise and S-wise galaxies showed that pattern recognition algorithms are able to predict the spin pattern of a galaxy by its photometric information in $\sim$64\% of the cases, significantly higher than mere chance of 50\% (P$<10^{-5}$). Similar results were obtained with 10,281 galaxies annotated automatically \citep{sha16}. These experiments showed that Z-wise and S-wise spiral galaxies in SDSS are not photometrically indifferent.

Experiments with a larger dataset of SDSS galaxies classified automatically allowed the identification of individual photometric measurements with statistically significant differences between Z-wise and S-wise galaxies. These experiments provided evidence that Z-wise and S-wise galaxies are different by their color \citep{shamir2013color,shamir2017colour}, while exhibiting strong statistical significance for difference in their brightness \citep{shamir2016photometric}. 

Since the experiments were done with galaxies acquired by one telescope (SDSS) and one image analysis method (Ganalyzer), it is possible that the results are driven by some unknown flaw in the telescope system, photometric pipeline, or automatic image analysis. This paper shows that the results are consistent across different telescopes and different image analysis methods, indicating the possibility that the reported results are not a reflection of a software or hardware flaw. The results also show that the asymmetry between Z-wise and S-wise galaxies changes with the direction of observation, and that opposite hemispheres have inverse asymmetry, suggesting that the asymmetry is aligned with a certain axis.

\section{DATA}
\label{computer_dataset}

Several datasets were used to show the asymmetry between galaxies with Z-wise and S-wise spin patterns, taken from two different sky surveys - the Sloan Digital Sky Survey, and the Panoramic Survey Telescope and Rapid Response System \citep{hodapp2004design,tonry2012pan}.  

An initial dataset of $3\cdot10^6$ SDSS DR8 galaxies with i magnitude lower than 18 and r Petrosian radius larger than 5.5'' was separated automatically into spiral and elliptical galaxies \citep{kum16}. Then, 740,908 galaxies identified as spiral were separated automatically into Z-wise and S-wise galaxies by applying the Ganalyzer algorithm \citep{shamir2011ganalyzer,ganalyzer_ascl,sha16}.

Ganalyzer works by first transforming the galaxy image to its radial intensity plot \citep{shamir2011ganalyzer}, which is a 360$\times$35 image that reflects the brightness variations around the center of the galaxy. The pixel (x,y) in the radial intensity plot is the median of the 5$\times$5 pixels around $(O_x+sin(\theta) \cdot r,O_y-cos(\theta) \cdot r)$ in the original galaxy image, where $(O_x,O_y)$ are the coordinates of the galaxy center, $\theta$ is the polar angle (in degrees), and r is the radial distance (in pixels), spanning over 35\% of the galaxy radius. That is, the x coordinate in the radial intensity plot corresponds to the polar angle, and the y coordinate corresponds to the radial distance from the galaxy center in the original image \citep{shamir2011ganalyzer}.

Then, each line in the radial intensity plot is searched for peaks, and the peaks are grouped by associating each detected peak $(x_0,y)$ in horizontal line y with the nearest peak $(x_1,y+1)$ in the y+1 horizontal line. The x coordinate of the two peaks are compared, and if $x_1<x_0$ the counter L is incremented, while if $x_1>x_0$ the counter R is incremented. If L$\geq$10 and L$\geq$3R then the galaxy is determined to have a Z-wise pattern, and if R$\geq$10 and R$\geq$3L the galaxy is determined to have a S-wise spin pattern. All other galaxies are assumed to have an undetermined spin pattern and are rejected from the analysis. Detailed description about the Ganalyzer algorithm is available in \citep{shamir2011ganalyzer,hoehn2014characteristics,shamir2012handedness}.

Applying the algorithm to the 740,908 spiral galaxies \citep{kum16} provided a dataset of 162,516 galaxies with identified spin patterns \citep{sha_input_2017,shamir2016photometric,shamir2017colour}. The spin patterns of the remaining galaxies were not identified (e.g., edge-on galaxies), and were therefore excluded from the experiment. The mean r magnitude of the galaxies in the dataset is $\sim$16.82 ($\sigma \simeq$0.99), and the mean redshift measured using 10,281 of the galaxies that had spectra is 0.086. 
The galaxies are in the declination range of (-11.3$^o$, 83.8$^o$). Manual inspection of 400 randomly selected Z-wise galaxies and 400 randomly selected S-wise galaxies showed that 24 galaxies classified as Z-wise and 21 galaxies classified as S-wise did not have identifiable spin patterns.

\begin{figure*}[ht]
\includegraphics[scale=0.48]{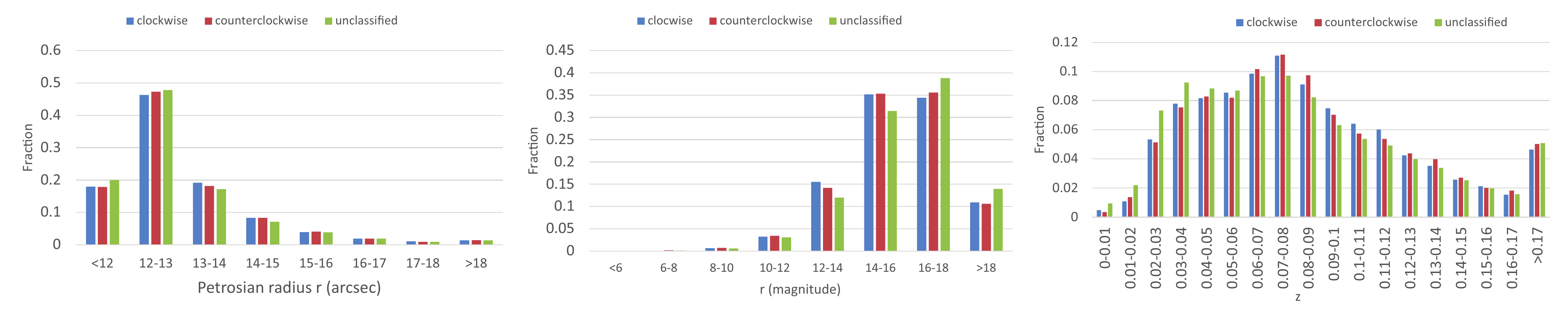}
\caption{Distribution of the r magnitude, Petrosian radius measured in the r band, and redshift among the dataset of automatically classified SDSS galaxies.}
\label{distribution}
\end{figure*}

Figure~\ref{distribution} displays the distribution of the Petrosian radius measured in the r band, the magnitude measured in the r band, and the redshift. Since most galaxies did not have spectra, the redshift is measured among a subset of 10,281 galaxies that had spectroscopic information. The figure shows the distribution of the galaxies with Z-wise patterns, S-wise patterns, and galaxies that were not assigned with an identifiable spin pattern and were excluded from the experiment.

The initial PanSTARRS data includes 3,053,831 celestial objects from data release 1 \citep{flewelling2016pan}, available at \footnote{https://panstarrs.stsci.edu}. That list includes 2,394,452 objects that were identified by PanSTARRS photometric pipeline \citep{chambers2016pan,tonry2012pan,magnier2016pan,waters2016pan} as extended sources in all bands. The rest of the galaxies were 659,379 objects not included in the first set, and had r Petrosian radius larger than 5.5'' and their PSF magnitude subtracted by their Kron magnitude was larger than 0.05. The brightness of all objects was limited to 19th magnitude in the r band.
 
\begin{figure*}[ht]
\includegraphics[scale=0.7]{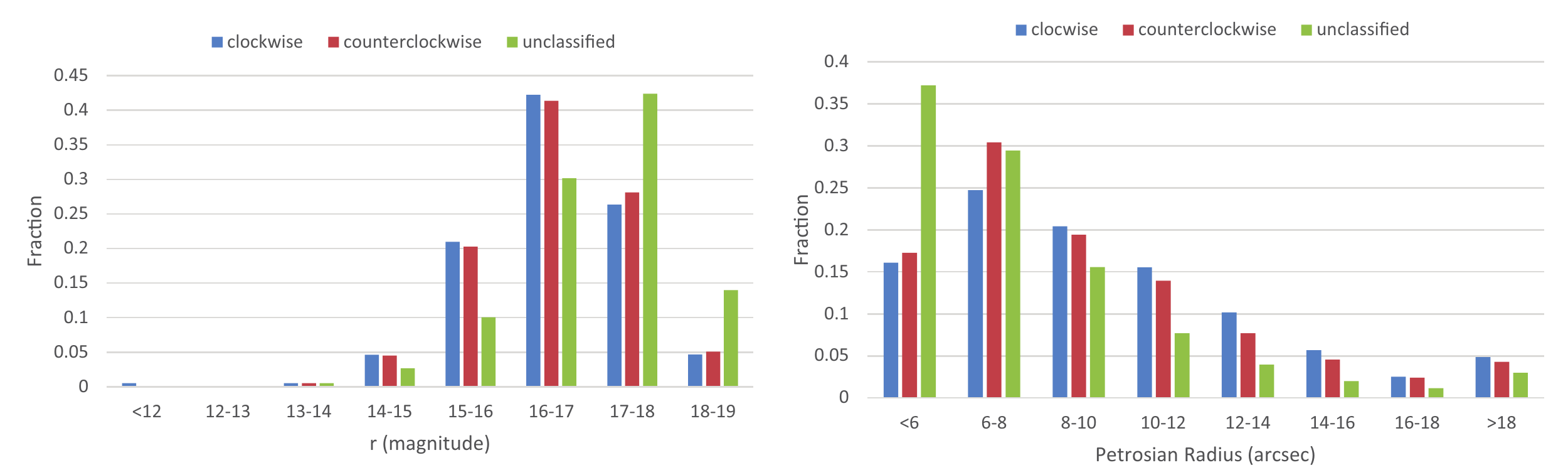}
\caption{Distribution of the r magnitude and Petrosian radius measured in the r band among the dataset of PanSTARRS galaxies.}
\label{distribution_PanSTARRS}
\end{figure*}

The images were downloaded as 120$\times$120 JPG images using the PanSTARRS {\it cutout} service in a process that required 62 days to complete. That dataset was reduced by the Ganalyzer algorithm \citep{shamir2011ganalyzer} into 29,013 galaxies annotated by their spin pattern. 
The remaining galaxies were not assigned with an identifiable spin pattern were excluded from the analysis. All galaxies were in the declination range of (-29.99$^o$, 84.94$^o$). Manual inspection of 200 galaxies showed that 11 galaxies did not have identifiable spin patterns. Figure~\ref{distribution_PanSTARRS} shows the r magnitude and Petrosian radius distribution of the galaxies in the PanSTARRS dataset. Only primary detections (rows with bestDetection=1) were used. For objects with multiple primary detections all detection were averaged. Values such as -9999 were ignored, as these values are flags and not actual photometric measurements.

\begin{figure*}[ht]
\includegraphics[scale=0.48]{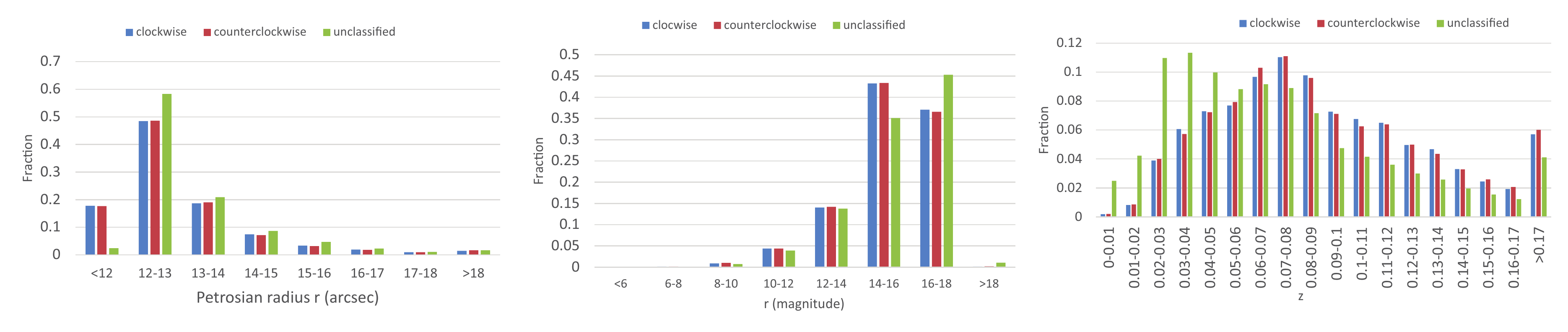}
\caption{Distribution of the r magnitude and Petrosian radius measured in the r band, and redshift in the dataset of manually annotated SDSS galaxies.}
\label{distribution_manual}
\end{figure*}

In addition to the automatically annotated datasets, 115,349 SDSS galaxies included in the catalog of SDSS spiral galaxies \citep{kum16} and also had spectroscopic information \citep{strauss2002spectroscopic} in SDSS DR8 were annotated manually. Since human classification of spin patterns can be substantially biased \citep{land2008galaxy,lintott2011galaxy}, the galaxies were mirrored randomly to balance a possible preference of a certain pattern, and the classification was done by the author in a careful process that required $\sim$450 hours of labor. The outcome of the process was a dataset of 20,712 galaxies with Z-wise spin patterns and 20,027 S-wise galaxies. The mean r magnitude of these galaxies is $\sim$16.6 ($\sigma \simeq$ 0.84), and the mean redshift is $\sim$0.09. 
Figure~\ref{distribution_manual} shows the distribution of the r magnitude, Petrosian radius, and redshift among the galaxies classified as Z, galaxies classified as S, and galaxies that their spin pattern could not be determined. 

The number of galaxies included in both the manually and the automatically classified SDSS datasets is 14,244. Examining the agreement between the two datasets showed that 6,881 galaxies classified manually as Z-wise were also classified automatically as Z-wise, while 276 galaxies classified manually as Z-wise were classified automatically as S-wise. Among the galaxies classified manually as S-wise 6,775 were classified automatically as S-wise, and 312 were classified by the automatic classifier as Z-wise.

The population of galaxies imaged by SDSS and PanSTARRS is not distributed uniformly in the sky. Table~\ref{sky_distribution} shows the number of galaxies in each of the datasets in each 30$^o$ RA range.

\begin{table*}[ht]
\caption{The distribution of the galaxy population in 30$^o$ RA ranges.}
\label{sky_distribution}
\begin{center}
{
 \small 
\begin{tabular}{lccc}
\hline \hline
RA           &    SDSS                             & SDSS                          & PanSTARRS  \\
range       &  (automatic annotation)      &  (manual annotation)     &                    \\   
\hline
0$^o$-30$^o$ & 25,418 & 1,229 & 3,113 \\
30$^o$-60$^o$ & 16,812 & 654 & 2,256 \\
60$^o$-90$^o$ & 3,250 & 18 & 1,349 \\
90$^o$-120$^o$ & 4,926 & 926 & 948 \\
120$^o$-150$^o$ & 19,396 & 6,907 & 2,868 \\
150$^o$-180$^o$ & 18,250 & 8,737 & 4,382\\
180$^o$-210$^o$ & 17,202 & 8,637 & 4,258 \\
210$^o$-240$^o$ & 17,539 & 8,632 & 3,890 \\
240$^o$-270$^o$ & 13,279 & 3,336 & 2,024 \\
270$^o$-300$^o$ & 809  & 3 & 441 \\
300$^o$-330$^o$ & 7,832  & 537 & 910 \\
330$^o$-360$^o$ & 17,803 & 1,123 & 2,574 \\
\hline
\end{tabular}
}
\end{center}
\end{table*}

\section{RESULTS}
\label{results}

The galaxy population is distributed such that the right ascension range of $(120^o, 210^o)$ contains 54,848, 24,281, and 11,508 galaxies in the automatically annotated SDSS dataset, the manually annotated SDSS dataset, and the PanSTARRS dataset, respectively. In the same RA range in the opposite hemisphere $(<30^o   ~\vee >300^o)$ the three datasets contained 51,053, 2,889, and 4,464 galaxies.

Tables~\ref{120_210} and~\ref{300_30} show the mean magnitude of Z-wise and S-wise galaxies in the $(120^o, 210^o)$ hemisphere and in the opposite $(<30^o  ~\vee >300^o)$ hemispheres, respectively. The tables show the exponential and de Vaucouleurs magnitudes, which are two magnitude models commonly used for galaxies, and are measured by both the SDSS and PanSTARRS photometric pipelines. 

\begin{table*}[ht]
\caption{Magnitude mean, standard error, and t-test of the difference between Z-wise and S-wise galaxies in the sky region of right ascension range of $(120^o, 210^o)$.}
\label{120_210}
\begin{center}
{
 \scriptsize 
\begin{tabular}{lccccc}
\hline \hline
Dataset & Magnitude &    Band    & Mean Z-wise & Mean S-wise & t-test  P  \\
            &  model      &                &                        &                                  &     \\   
\hline
SDSS automatically annotated & de Vaucouleurs & u & 18.205$\pm$0.007 & 18.178$\pm$0.007 & 0.004 \\
SDSS automatically annotated & de Vaucouleurs & g & 17.047$\pm$0.006 & 17.027$\pm$0.006 & 0.012 \\
SDSS automatically annotated & de Vaucouleurs & r & 16.455$\pm$0.006 & 16.434$\pm$0.006 & 0.008 \\
SDSS automatically annotated & de Vaucouleurs & i & 16.138$\pm$0.006 & 16.119$\pm$0.006 & 0.028 \\
SDSS automatically annotated & de Vaucouleurs & z & 15.939$\pm$0.006 & 15.919$\pm$0.006 & 0.027 \\
SDSS automatically annotated & Exponential & u & 18.782$\pm$0.006 & 18.757$\pm$0.006 & 0.004 \\
SDSS automatically annotated & Exponential & g & 17.503$\pm$0.006 & 17.482$\pm$0.006 & 0.016 \\
SDSS automatically annotated & Exponential & r & 16.913$\pm$0.006 & 16.892$\pm$0.006 & 0.008 \\
SDSS automatically annotated & Exponential & i & 16.597$\pm$0.006 & 16.578$\pm$0.006 & 0.021 \\
SDSS automatically annotated & Exponential & z & 16.435$\pm$0.006 & 16.416$\pm$0.006 & 0.033 \\
SDSS manually annotated & de Vaucouleurs & u & 17.941$\pm$0.008 & 17.919$\pm$0.008 & 0.059 \\
SDSS manually annotated & de Vaucouleurs & g & 16.818$\pm$0.007 & 16.795$\pm$0.007 & 0.027 \\
SDSS manually annotated & de Vaucouleurs & r & 16.229$\pm$0.007 & 16.206$\pm$0.007 & 0.021 \\
SDSS manually annotated & de Vaucouleurs & i & 15.897$\pm$0.007 & 15.873$\pm$0.007 & 0.023 \\
SDSS manually annotated & de Vaucouleurs & z & 15.675$\pm$0.008 & 15.643$\pm$0.008 & 0.003 \\
SDSS manually annotated & Exponential & u & 18.551$\pm$0.008 & 18.526$\pm$0.008 & 0.033 \\
SDSS manually annotated & Exponential & g & 17.273$\pm$0.007 & 17.247$\pm$0.008 & 0.022 \\
SDSS manually annotated & Exponential & r & 16.683$\pm$0.007 & 16.657$\pm$0.008 & 0.013 \\
SDSS manually annotated & Exponential & i & 16.359$\pm$0.007 & 16.333$\pm$0.008 & 0.014 \\
SDSS manually annotated & Exponential & z & 16.193$\pm$0.008 & 16.161$\pm$0.008 & 0.003 \\

PanSTARRS & de Vaucouleurs & g & 17.028$\pm$0.02 & 16.934$\pm$0.02 & $2\cdot 10^{-5}$ \\
PanSTARRS & de Vaucouleurs & r & 16.204$\pm$0.01 & 16.142$\pm$0.01 & 0.0001 \\
PanSTARRS & de Vaucouleurs & i & 16.171$\pm$0.01 & 16.11$\pm$0.01 & $4\cdot 10^{-5}$ \\
PanSTARRS & de Vaucouleurs & z & 15.781$\pm$0.01 & 15.736$\pm$0.01 & 0.003 \\
PanSTARRS & de Vaucouleurs & y & 15.893$\pm$0.01 & 15.841$\pm$0.01 & 0.0006 \\

PanSTARRS & Exponential & g & 17.054$\pm$0.01 & 16.986$\pm$0.01 & $2.8\cdot10^{-5}$ \\
PanSTARRS & Exponential & r & 16.538$\pm$0.01 & 16.471$\pm$0.01 & $1.5\cdot10^{-5}$ \\
PanSTARRS & Exponential & i & 16.236$\pm$0.01 & 16.171$\pm$0.01 & $1.2\cdot10^{-5}$ \\
PanSTARRS & Exponential & z & 16.106$\pm$0.01 & 16.038$\pm$0.01 & $4.6\cdot10^{-6}$ \\
PanSTARRS & Exponential & y & 15.931$\pm$0.01 & 15.897$\pm$0.01 & $7.3\cdot10^{-6}$ \\


\hline
\end{tabular}
}
\end{center}
\end{table*}

\begin{table*}[ht]
\caption{Magnitude mean, standard error, and t-test of the difference between Z-wise and S-wise galaxies in the region of $(<30^o ~\vee >300^o)$.}
\label{300_30}
\begin{center}
{
 \scriptsize
\begin{tabular}{lccccc}
\hline \hline
Dataset & Magnitude &    Band    & Mean Z-wise & Mean S-wise & t-test  P  \\
            &  model      &                &                        &                                  &  (two tails)  \\   
\hline
SDSS automatically annotated & de Vaucouleurs & u & 18.245$\pm$0.007 & 18.301$\pm$0.007 & 4.6$\cdot$10-8 \\
SDSS automatically annotated & de Vaucouleurs & g & 17.035$\pm$0.006 & 17.089$\pm$0.006 & 2.9$\cdot$10-10 \\
SDSS automatically annotated & de Vaucouleurs & r & 16.409$\pm$0.006 & 16.459$\pm$0.006 & 1.3$\cdot$10-9 \\
SDSS automatically annotated & de Vaucouleurs & i & 16.076$\pm$0.006 & 16.129$\pm$0.006 & 6.5$\cdot$10-10 \\
SDSS automatically annotated & de Vaucouleurs & z & 15.856$\pm$0.007 & 15.912$\pm$0.007 & 1.9$\cdot$10-9 \\
SDSS automatically annotated & Exponential & u & 18.83$\pm$0.007 & 18.883$\pm$0.007 & 3.3$\cdot$10-8 \\
SDSS automatically annotated & Exponential & g & 17.508$\pm$0.006 & 17.564$\pm$0.006 & 1.3$\cdot$10-10 \\
SDSS automatically annotated & Exponential & r & 16.886$\pm$0.006 & 16.937$\pm$0.006 & 2.1$\cdot$10-9 \\
SDSS automatically annotated & Exponential & i & 16.549$\pm$0.006 & 16.601$\pm$0.006 & 2$\cdot$10-9 \\
SDSS automatically annotated & Exponential & z & 16.36$\pm$0.006 & 16.415$\pm$0.006 & 2.43$\cdot$10-9 \\
SDSS manually annotated & de Vaucouleurs & u & 18.013$\pm$0.02 & 18.015$\pm$0.02 & 0.93 \\
SDSS manually annotated & de Vaucouleurs & g & 16.855$\pm$0.02 & 16.861$\pm$0.02 & 0.86 \\
SDSS manually annotated & de Vaucouleurs & r & 16.241$\pm$0.02 & 16.261$\pm$0.02 & 0.51 \\
SDSS manually annotated & de Vaucouleurs & i & 15.893$\pm$0.02 & 15.918$\pm$0.02 & 0.4 \\
SDSS manually annotated & de Vaucouleurs & z & 15.664$\pm$0.02 & 15.689$\pm$0.02 & 0.44 \\
SDSS manually annotated & Exponential & u & 18.639 & 18.648 & 0.8 \\
SDSS manually annotated & Exponential & g & 17.338 & 17.347 & 0.76 \\
SDSS manually annotated & Exponential & r & 16.718 & 16.738 & 0.52 \\
SDSS manually annotated & Exponential & i & 16.374 & 16.398 & 0.44 \\
SDSS manually annotated & Exponential & z & 16.193 & 16.219 & 0.43 \\

PanSTARRS & de Vaucouleurs & g & 16.963$\pm$0.02 & 17.058$\pm$0.03 & 0.001 \\
PanSTARRS & de Vaucouleurs & r & 16.234$\pm$0.02 & 16.269$\pm$0.01 & 0.08 \\
PanSTARRS & de Vaucouleurs & i & 16.223$\pm$0.01 & 16.248$\pm$0.01 & 0.075 \\
PanSTARRS & de Vaucouleurs & z & 16.278$\pm$0.01 & 16.297$\pm$0.01 & 0.177 \\
PanSTARRS & de Vaucouleurs & y & 15.867$\pm$0.01 & 15.890$\pm$0.01 & 0.113 \\

PanSTARRS & Exponential & g & 17.039$\pm$0.01 & 17.072$\pm$0.01 & 0.085 \\
PanSTARRS & Exponential & r & 16.521$\pm$0.01 & 16.55$\pm$0.01 & 0.092 \\
PanSTARRS & Exponential & i & 16.195$\pm$0.01 & 16.222$\pm$0.01 & 0.097 \\
PanSTARRS & Exponential & z & 16.058$\pm$0.01 & 16.081$\pm$0.01 & 0.187 \\
PanSTARRS & Exponential & y & 15.87$\pm$0.01 & 15.896$\pm$0.01 & 0.125 \\


\hline
\end{tabular}
}
\end{center}
\end{table*}

While it has been shown that Z-wise and S-wise galaxies imaged by SDSS are photometrically different from each other \citep{shamir2013color,sha16,shamir2017colour,shamir2016photometric}, Tables~\ref{120_210} and~\ref{300_30} show that the asymmetry is consistent across hemispheres, so that the asymmetry between Z-wise and S-wise galaxies in $(120^o, 210^o)$ is inverse to the asymmetry in the same RA range in the opposite hemisphere $(<30^o ~\vee >300^o)$. Other magnitude models (PSF magnitude, Kron magnitude, Petrosian magnitude) were tested and showed the same differences, indicating that the asymmetry is not the result of an error in a certain magnitude model. 

All three datasets show with strong statistical power that galaxies with S-wise spin pattern in RA range $(120^o, 210^o)$ are brighter than galaxies with Z-wise spin patterns. The RA range of $(<30^o ~\vee >300^o)$ shows asymmetry with strong statistical significance observed in the automatically annotated SDSS galaxies. The manually annotated SDSS galaxies do not show statistically significant asymmetry in that RA range, possibly due to the lower number of galaxies, but it is aligned with the automatically annotated SDSS galaxies, showing that Z-wise galaxies in that dataset are brighter in that RA range compared to galaxies with S-wise patterns. The difference in the PanSTARRS galaxies in that RA range is statistically significant when using the one-tailed distribution, given the hypothesis that Z-wise galaxies in that sky region are brighter, as shown by the automatically annotated galaxies. For instance, the chance probability that Z-wise galaxies are brighter in the r exponential magnitude in both the manually annotated SDSS galaxies and the PanSTARRS dataset in the RA range $(<30^o ~\vee >300^o)$ is $\simeq0.025$.

Repeating the experiment after mirroring half of the galaxies randomly provided no statistically significant difference in any of the datasets. The dataset of manually annotated galaxies had spectroscopic information, showing no statistical significance between the mean redshift of the galaxies. The mean redshift of the Z-wise and S-wise galaxies in the RA range of $(120^o, 210^o)$ is 0.089$\pm$0.002 and 0.087$\pm$0.002, respectively, and it is 0.082$\pm$0.002 and 0.083$\pm$0.002 for Z-wise and S-wise galaxies in the opposite hemisphere $(<30^o ~\vee >300^o)$, showing no statistically significant difference in the redshift.

\begin{figure}[h!]
\includegraphics[scale=0.65]{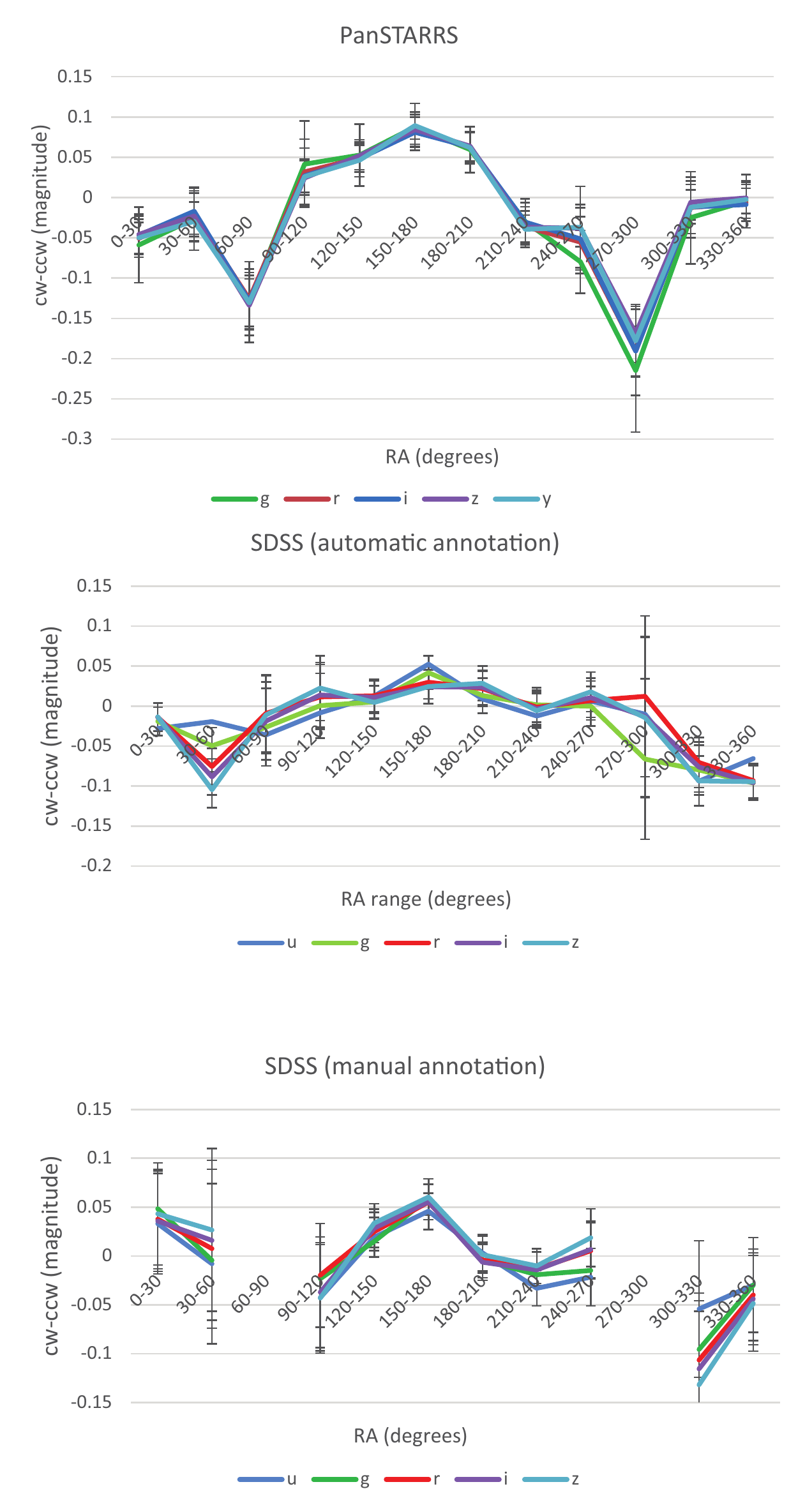}
\caption{Asymmetry between the exponential magnitude of Z-wise and S-wise galaxies in different RA ranges using SDSS automatically annotated galaxies, SDSS manually annotated galaxies, and PanSTARRS galaxies. The error bars show the standard error of the mean.}
\label{asymmetry_exp}
\end{figure}

\begin{figure}[h!]
\includegraphics[scale=0.65]{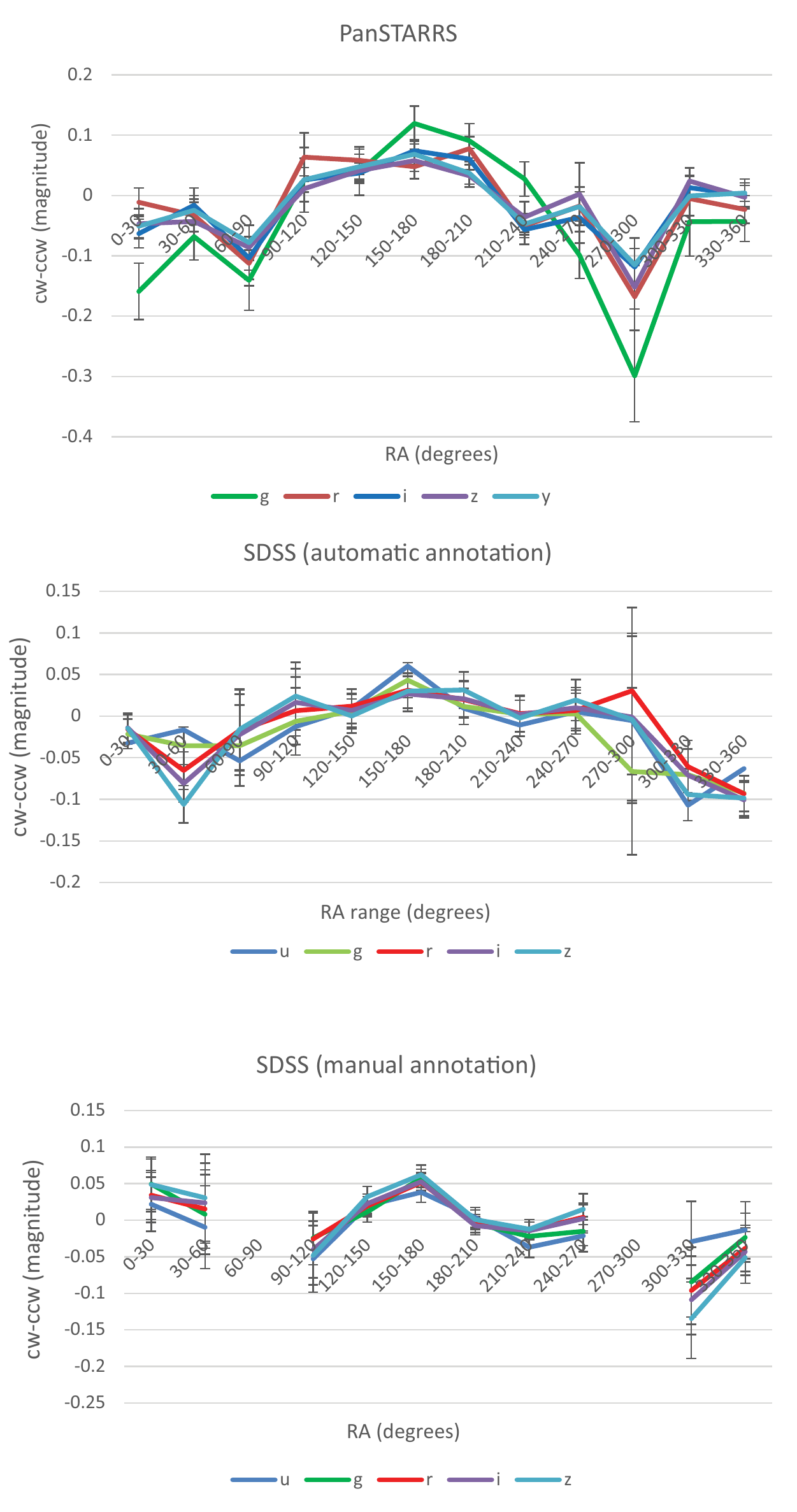}
\caption{Asymmetry between the de Vaucouleurs magnitude of Z-wise and S-wise galaxies in different RA ranges using SDSS automatically annotated galaxies, SDSS manually annotated galaxies, and PanSTARRS galaxies.}
\label{asymmetry_dev}
\end{figure}

Figures~\ref{asymmetry_exp} and~\ref{asymmetry_dev} show the asymmetry measured using the exponential magnitude and the de Vaucouleurs magnitude, respectively, in the different right ascension ranges. The right ascension ranges of $(270^o,300^o)$ and $(60^o,90^o)$ in the SDSS manually annotated galaxies were not used due to the low number of galaxies in these ranges, 1 and 20, respectively. The figures show that the asymmetry changes with the right ascension also inside the hemispheres, and that the difference between the brightness of S-wise galaxies and Z-wise galaxies changes with the right ascension.

\begin{figure*}[h!]
\includegraphics[scale=0.7]{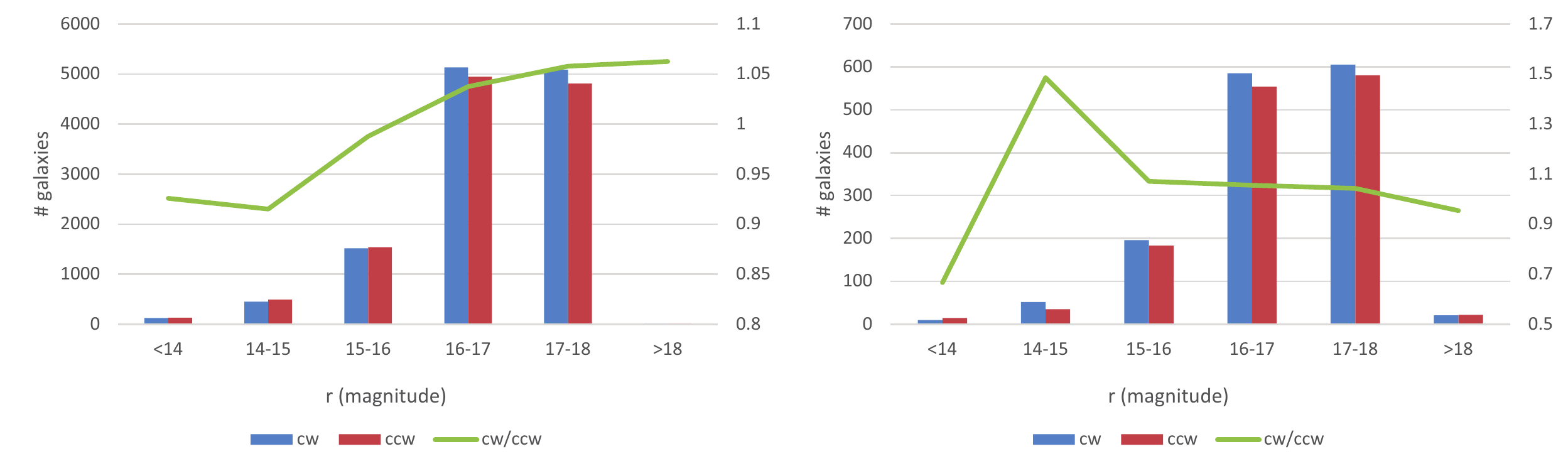}
\caption{Histogram of the manually classified galaxies in the different exponential r magnitude ranges. The figure shows the distribution in the $(120^o, 210^o)$ RA range (left) and $(<30^o \vee >300^o)$ (right).}
\label{manual_r_exp_histogram}
\end{figure*}

\begin{figure*}[h!]
\includegraphics[scale=0.7]{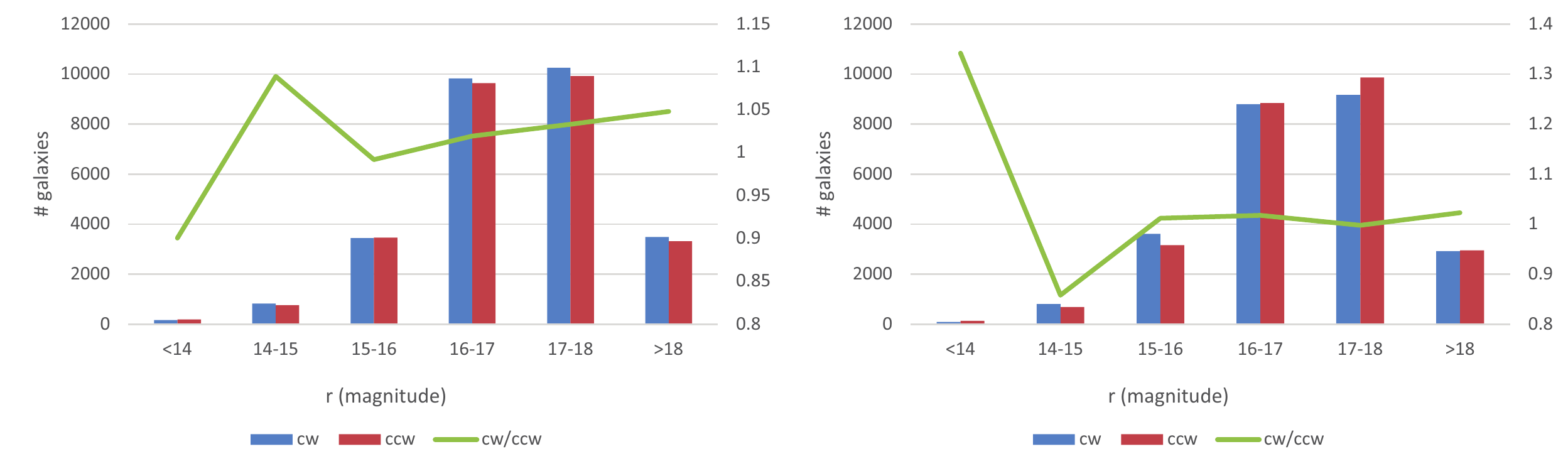}
\caption{Histogram of the automatically classified SDSS galaxies in the different exponential r magnitude ranges. The figure shows the distribution in the $(120^o, 210^o)$ RA range (left) and the $(<30^o \vee >300^o)$ range (right).}
\label{automatic_r_exp_histogram}
\end{figure*}

\begin{figure*}[h!]
\includegraphics[scale=0.7]{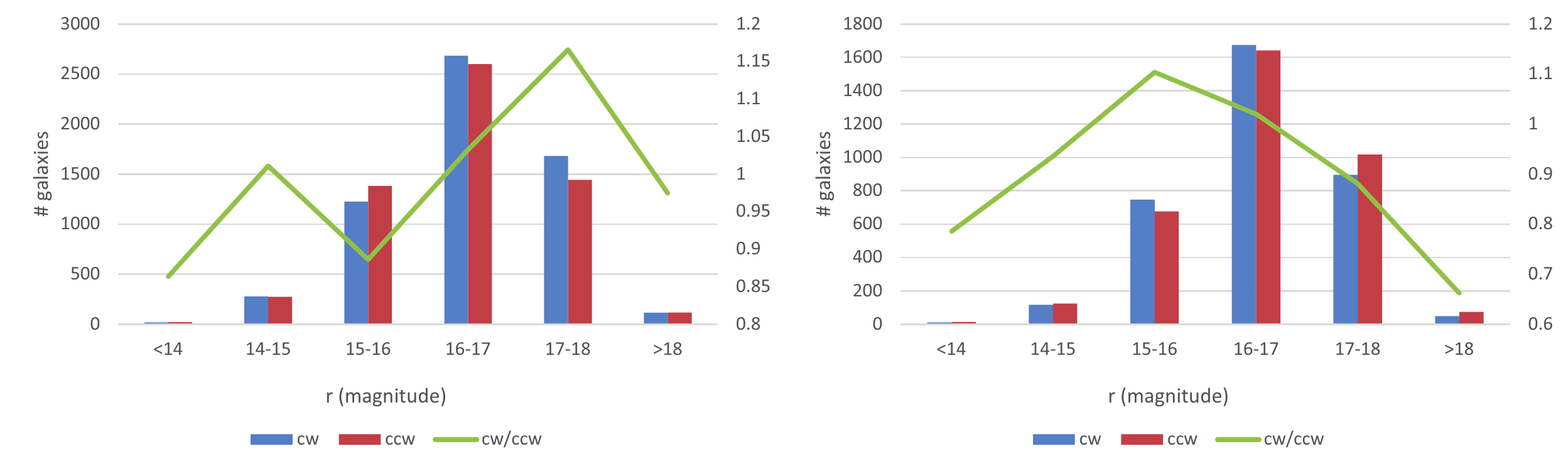}
\caption{Histogram of the PanSTARRS galaxies in the different exponential r magnitude ranges. The graph shows the distribution in the $(120^o, 210^o)$ RA range (left) and the $(<30^o \vee >300^o)$ range (right).}
\label{PanSTARRS_r_exp_histogram}
\end{figure*}

The tables and graphs show that the asymmetry is strongest at around the RA range of $(120^o, 210^o)$, and the same RA range in the opposite hemisphere $(<30^o ~\vee >300^o)$. Figures~\ref{manual_r_exp_histogram},~\ref{automatic_r_exp_histogram}, and~\ref{PanSTARRS_r_exp_histogram} show the histograms of the distribution of the galaxies in different exponential r magnitude ranges in the $(120^o, 210^o)$ and  $(<30^o ~\vee >300^o)$ RA ranges.

\subsection{Spin asymmetry in Galaxy Zoo 1 data}
\label{gz1}

Galaxy Zoo \citep{lintott2011galaxy} is a citizen science campaign that aimed at classifying SDSS galaxies by their morphology, and also collected manual citizen science annotations of Z-wise and S-wise galaxies. Although the manual annotation of Z-wise and S-wise galaxies was found to be heavily biased \citep{hayes2016nature}, a subset of galaxies classified by the citizen scientists with a high degree of 95\% ``superclean''  \citep{land2008galaxy} can be used to test whether the different magnitude between Z-wise and S-wise galaxies is also consistent when using galaxies annotated manually as part of Galaxy Zoo. 

Comparing the Galaxy Zoo galaxies to the dataset described in Section~\ref{computer_dataset} showed that 8,710 galaxies were included in both datasets (4,133 Z-wise galaxies and 4,577 S-wise galaxies). Of these galaxies, 1,199 of the galaxies classified as Z-wise in Galaxy Zoo were also classified as Z-wise in the automatically classified dataset, and 58 galaxies classified as Z-wise were classified automatically as S-wise. Among the galaxies classified by Galaxy Zoo as S-wise 1,292 were classified automatically as S-wise and 66 as Z-wise.

When comparing to the dataset of manually classified galaxies, 2,045 Z-wise galaxies were also classified as Z-wise in the manually classified galaxies, and 14 galaxies classified as Z-wise were classified as S-wise. Among galaxies classified in the Galaxy Zoo 1 dataset as S-wise 2,287 were classified as S-wise also in the dataset of manually classified galaxies, and 20 as Z-wise.

Table~\ref{galaxy_zoo} shows the magnitude of Z-wise and S-wise galaxies measured with 5,923 galaxies (2,804 Z-wise and 3,119 S-wise) within the RA range of $(120^o, 210^o)$ and had an agreement of more than 95\% of the votes. 

\begin{table*}[ht]
\caption{Magnitude mean, standard error, and t-test of the difference between the means of the ``superclean'' Z-wise and S-wise galaxies annotated by Galaxy Zoo in the right ascension range of $(120^o, 210^o)$.}
\label{galaxy_zoo}
\begin{center}
{
 \small 
\begin{tabular}{lcccc}
\hline \hline
Magnitude &    Band    & Mean Z-wise & Mean S-wise & t-test  P  \\
model       &                &                        &                     &                \\   
\hline
de Vaucouleurs & u & 17.094$\pm$0.02 & 17.095$\pm$0.02 & 0.97 \\
de Vaucouleurs & g & 15.927$\pm$0.02 & 15.923$\pm$0.02 & 0.88 \\
de Vaucouleurs & r & 15.373$\pm$0.02 & 15.356$\pm$0.02 & 0.54 \\
de Vaucouleurs & i & 15.059$\pm$0.02 & 15.033$\pm$0.02 & 0.36 \\
de Vaucouleurs & z & 14.846$\pm$0.02 & 14.814$\pm$0.02 & 0.26 \\
Exponential & u & 17.656$\pm$0.02 & 17.658$\pm$0.02 & 0.94 \\
Exponential & g & 16.351$\pm$0.02 & 16.343$\pm$0.02 & 0.77 \\
Exponential & r & 15.79$\pm$0.02 & 15.771$\pm$0.02 & 0.5 \\
Exponential & i & 15.492$\pm$0.02 & 15.467$\pm$0.02 & 0.38 \\
Exponential & z & 15.373$\pm$0.02 & 15.339$\pm$0.02 & 0.23 \\
\hline
\end{tabular}
}
\end{center}
\end{table*}



The table shows that the S-wise galaxies annotated by Galaxy Zoo are also generally brighter than the Z-wise galaxies, which is aligned with the asymmetry observed in the other datasets and shown in Tables~\ref{120_210} and~\ref{300_30}. The differences are not statistically significant, with the highest t-test statistical significance of $\sim$0.23 observed in the exponential Z-wise magnitude. It should be noted that the number of Galaxy Zoo galaxies with ``superclean'' annotations is much lower than the number of galaxies in the other datasets, and therefore statistically significant difference is not expected. The same RA range in the opposite hemisphere had a much lower number of just 687 galaxies with ``superclean'' agreement between the voters, and the low number does not allow a meaningful analysis.


\subsection{Spin asymmetry in Galaxy Zoo 2 data}
\label{gz2}

Galaxy Zoo 2 \citep{willett2013galaxy} provided annotations of the morphology of SDSS galaxies, and a dataset of 13,440 manually annotated Galaxy Zoo 2 galaxies \citep{sha_input_2016} was used in to show that pattern recognition algorithms can identify between Z-wise and S-wise galaxies with accuracy significantly higher than mere chance \citep{sha16}. The RA range $(120^o,210^o)$ contained 6,974 galaxies. Table~\ref{galaxy_zoo2} shows the magnitude of Z-wise and S-wise Galaxy Zoo 2 galaxies. As the table shows, Galaxy Zoo 2 galaxies show that S-wise galaxies in that RA range are brighter than Z-wise galaxies, and these results are aligned with the observations of the SDSS automatically classified galaxies and PanSTARRS. These results agree with the automatically annotated galaxies, but the statistical power of the difference is not expected to be high due to the much lower number of galaxies. The RA range $(<30^o ~\vee >300^o)$ in that dataset contained no galaxies.

\begin{table*}[ht]
\caption{Magnitude mean, standard error, and t-test of the difference between the means of Z-wise and S-wise Galaxy Zoo 2 galaxies in the right ascension range of $(120^o, 210^o)$.}
\label{galaxy_zoo2}
\begin{center}
{
 \small 
\begin{tabular}{lcccc}
\hline \hline
Magnitude &    Band    & Mean Z-wise & Mean S-wise & t-test  P  \\
model       &                &                        &                                  &     \\   
\hline
de Vaucouleurs & u & 17.085$\pm$0.02 & 17.033$\pm$0.02 & 0.025 \\
de Vaucouleurs & g & 15.896$\pm$0.01 & 15.870$\pm$0.02 & 0.24 \\
de Vaucouleurs & r & 15.351$\pm$0.01 & 15.318$\pm$0.02 & 0.14 \\
de Vaucouleurs & i & 15.024$\pm$0.02 & 14.988$\pm$0.02 & 0.108 \\
de Vaucouleurs & z & 14.827$\pm$0.02 & 14.775$\pm$0.02 & 0.019\\
Exponential & u & 17.674$\pm$0.02 & 17.615$\pm$0.02 & 0.014 \\
Exponential & g & 16.318$\pm$0.02 & 16.280$\pm$0.02 & 0.107 \\
Exponential & r & 15.760$\pm$0.02 & 15.718$\pm$0.02 & 0.071 \\
Exponential & i & 15.457$\pm$0.02 & 15.411$\pm$0.02 & 0.056 \\
Exponential & z & 15.347$\pm$0.02 & 15.293$\pm$0.02 & 0.025 \\
\hline
\end{tabular}
}
\end{center}
\end{table*}

Comparing the Galaxy Zoo 2 galaxies to the galaxies described in Section~\ref{computer_dataset} showed that 8,710 galaxies were included in both datasets (5,788 Z-wise galaxies and 5,328 S-wise galaxies). Of these galaxies, 1,736 of the galaxies classified as Z-wise in the Galaxy Zoo 2 dataset were also classified as Z-wise in the automatically classified dataset, and 87 galaxies classified as Z-wise were classified automatically as S-wise. Among the galaxies classified by Galaxy Zoo 2 as S-wise 1,746 were classified automatically as S-wise and 70 as Z-wise.

When comparing to the manually classified galaxies, 2,703 Z-wise galaxies were also classified as Z-wise in the manually classified galaxies, and 41 as S-wise. Among galaxies classified in the Galaxy Zoo 2 dataset as S-wise 2,643 were classified as S-wise, and 44 as Z-wise.

\subsection{Redshift dependence}
\label{redshift}

\begin{figure*}[ht]
\includegraphics[scale=0.7]{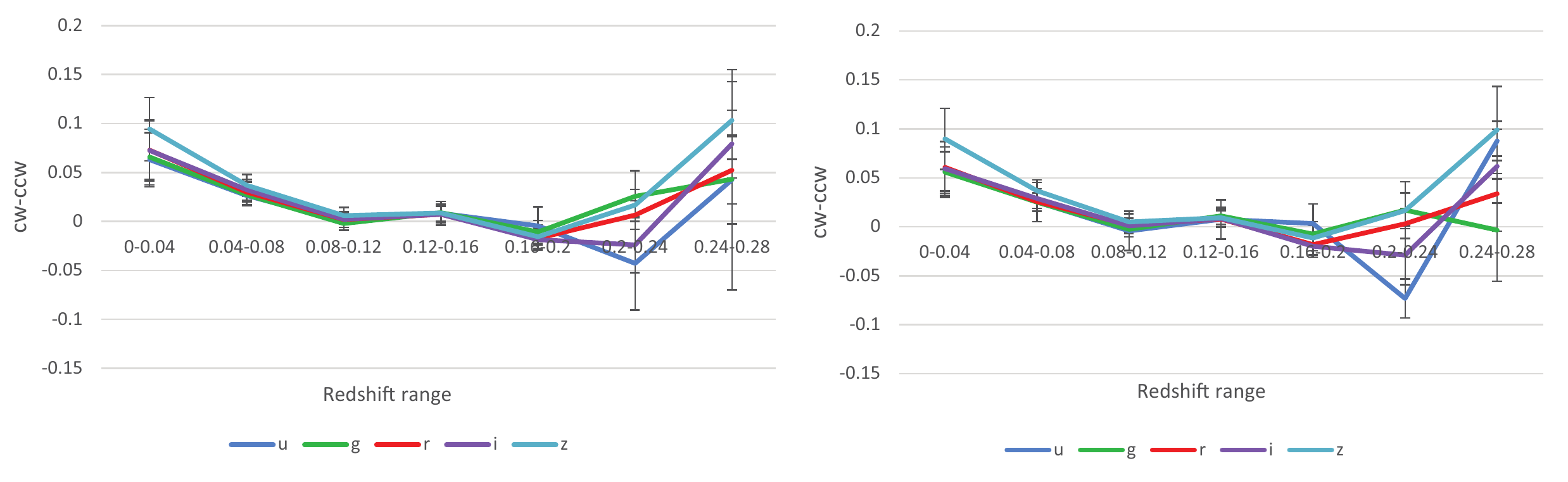}
\caption{The change in asymmetry in the exponential magnitude (left) and the de Vaucouleurs magnitude (right) in different redshift ranges.}
\label{z_ranges}
\end{figure*}

The galaxies in the manually annotated dataset have spectra, allowing to profile the change in the asymmetry in different redshift ranges. Figure~\ref{z_ranges} shows the asymmetry measured using the exponential and de Vaucouleurs magnitudes in different redshift ranges in the RA range of $(120^o,210^o)$. The figure shows that the asymmetry decreases when the redshift gets higher, which can also be attributed to higher inaccuracy of the classification of more distant objects. The asymmetry increases when the redshift is higher than 0.2, but the increase is not statistically significant. Photometric redshift cannot be used as a valid measurement as it depends on the magnitude itself, so magnitude differences between Z-wise and S-wise galaxies would also affect the photometric redshift. 

\subsection{A possible spin asymmetry axis}
\label{axis}


The opposite asymmetry in opposite hemispheres indicates that the asymmetry could be aligned with an axis, such that galaxies with spin patterns oriented around that axis are brighter than galaxies with opposite spin patterns. To identify the approximate location of such possible axis, the SDSS automatically and manually annotated galaxies were combined into a single dataset, and the sky was separated into regions with half-width size of 10$^o$, such that no two regions had overlapping parts of the sky. To avoid underpopulated areas, only regions with 3,000 galaxies or more were used. That provided a set of 28 non-overlapping sky regions, and the asymmetry in each sky region was computed.

\begin{figure}[h!]

\includegraphics[scale=0.65]{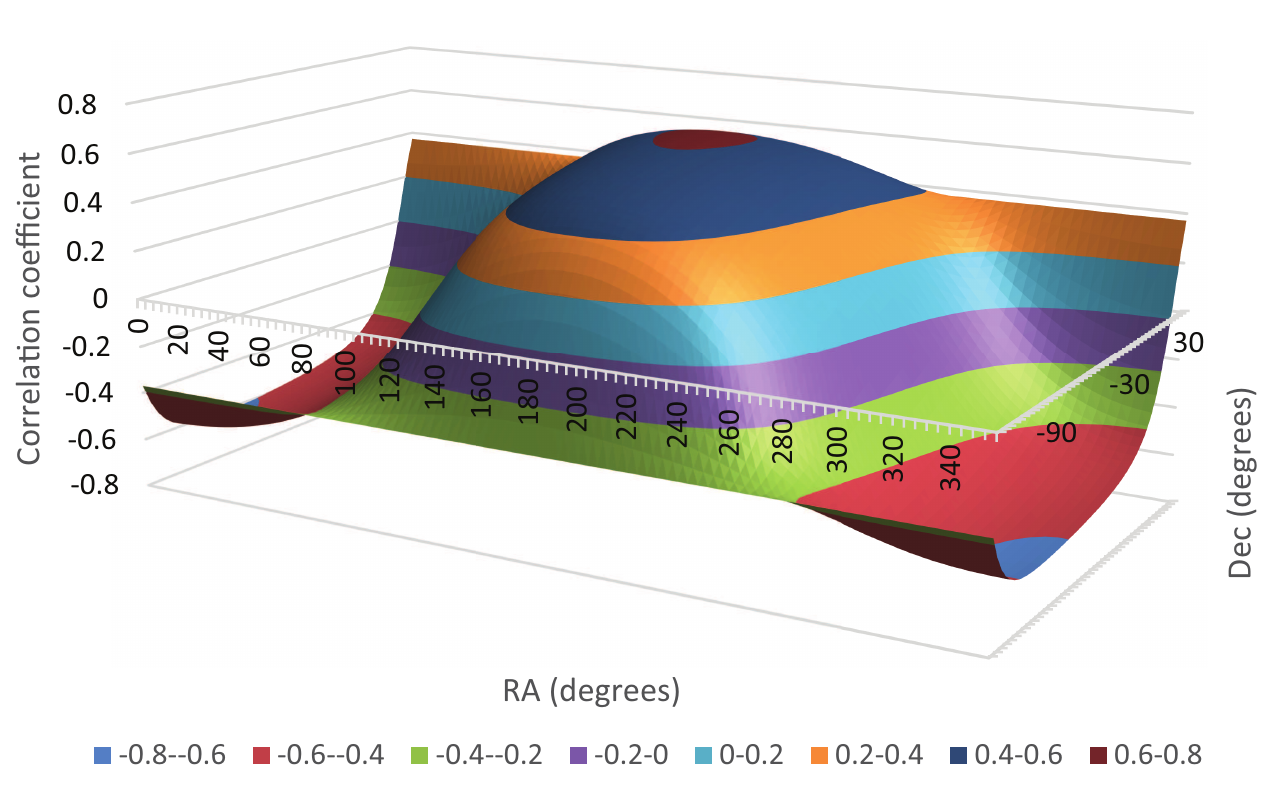}
\caption{The Pearson correlation between the exponential magnitude asymmetry and the cosine of the angular distance from all possible integer ($\alpha$,$\delta$).}
\label{correlation_exp}
\end{figure}

Then, all possible integer right ascension and declination combinations ($\alpha$,$\delta$) were tested for cosine dependence by computing the Pearson correlation between vectors D and A, such that Di is the cosine of the angular distances between ($\alpha$,$\delta$) and the center of region i, and Ai is the difference between the mean exponential magnitude of the Z-wise galaxies and the mean exponential magnitude of the S-wise galaxies in region i. That was repeated for all five bands, and the correlation of each ($\alpha$,$\delta$) is the mean correlation measured using all bands. Figure~\ref{correlation_exp} shows the Pearson correlation coefficient measured from all possible integer ($\alpha$,$\delta$).

The highest Pearson correlation coefficient was $\sim$0.61 observed at ($\alpha$=172$^o$,$\delta$=50$^o$). Repeating the analysis 1,000 times with randomly assigned galaxy spin patterns and selecting the maximum Pearson correlation in each run provided a mean of $\sim$0.28 and standard deviation of $\sim$0.11 (z-score of $\sim$0.41). The correlation coefficient with the z-score of 0.41 from 0.61 is $\sim$0.53, and therefore the 1$\sigma$ error range of the right ascension is (132$^o$,224$^o$), and for the declination it is (-26$^o$,74$^o$). It should be noted that the most probable axis of ($\alpha$=172$^o$,$\delta$=50$^o$) is roughly aligned with the galactic pole ($\alpha$=192$^o$, $\delta$=27$^o$). The right ascension is also close to the right ascension of the the CMB dipole \citep{aghanim2014planck} at 166$^o$, but there is substantial difference in the declination (50$^o$ compared to -27$^o$).



\begin{figure}[h!]
\includegraphics[scale=0.65]{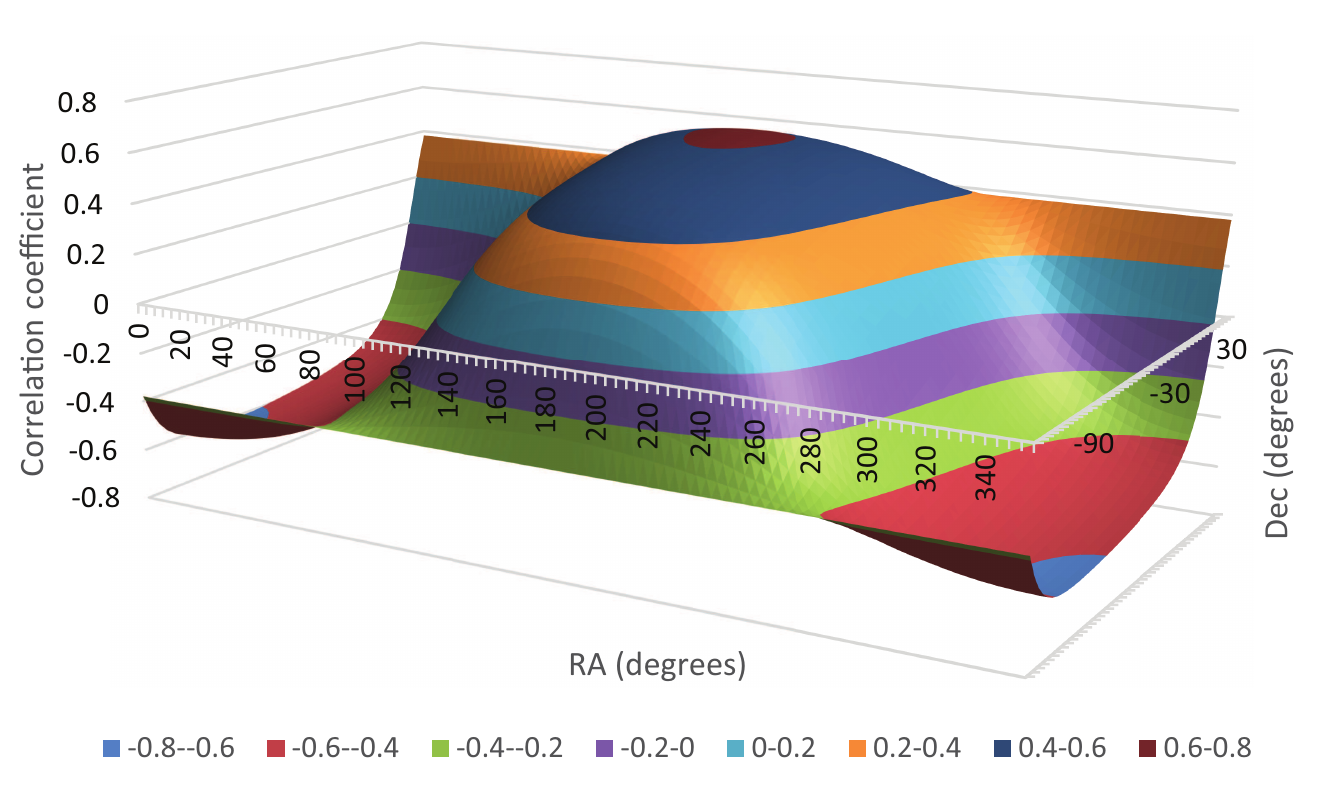}
\caption{The Pearson correlation between the de Vaucouleurs magnitude asymmetry and the cosine of the angular distance from all possible integer ($\alpha$,$\delta$).}
\label{correlation_dev}
\end{figure}

The same experiment was repeated with the de Vaucouleurs, and Figure~\ref{correlation_dev} shows the correlation coefficient at each possible integer right ascension and declination combinations. The highest Pearson correlation of $\sim$0.61 is observed at ($\alpha$=175$^o$,$\delta$=50$^o$), which is in close agreement with the exponential magnitude.


\section{DISCUSSION}
\label{discussion}

The spin pattern of a spiral galaxy is a crude morphological feature that is sensitive to the perspective of the observer. It is therefore expected that in a sufficiently large population of galaxies the physical properties of galaxies with Z-wise spin patterns will not be different from the physical properties of galaxies with S-wise spin patterns. The observations presented in this paper show the existence of small but significant asymmetry between galaxies with opposite spin patterns. Moreover, the asymmetry in the differences between galaxies with opposite spin patterns is inverse in opposite hemispheres. That is, galaxies with S-wise spin patterns are significantly brighter in the RA range of $(120^o,210^o)$, but are significantly dimmer in the same RA range in the opposite hemisphere of $(<30^o ~\vee >300^o)$. That inverse asymmetry can suggest the existence of an axis around which galaxy spin seem brighter to an Earth-based observer. Clearly, with mean redshift of $\sim$0.09 the region analyzed in this study is far larger than the size of a supercluster or any other known structure. 
 
While the observation could be related to incompleteness of current cosmological theories \citep{kroupa2012dark}, the proximity of the most probable axis to the galactic pole might suggest that the asymmetry could also be the effect of solar motion, and the asymmetry can be explained by relativistic beaming \citep{loeb2003periodic}. With relativistic beaming, stars in a host galaxy that rotate in the same direction of the Sun would seem to an Earth-based observer brighter than stars in a host galaxy that spins in the opposite direction. Since the difference of the bolometric flux is by a fraction of $4\cdot\frac{v}{c}$ \citep{loeb2003periodic}, assuming $\frac{v}{c}$ of the Sun of $\sim$0.001 and a similar motion of the star in the observed galaxy, the maximum asymmetry can be $\sim$0.8\%. Since the asymmetry observed in the statistically significant galaxy population is not greater than 0.6\%, and it peaks at around the galactic pole, it is possible that the observed asymmetry is a large scale observation of relativistic beaming. 

The difference in the velocity between the Sun and another Sun-like star in another face-on galaxy rotating in the opposite direction is $2 \cdot 0.001 \cdot c$, or $z \simeq 0.002$. The K corrections for the different filters in that redshift are approximately 0.01, -0.07, -0.02, 0, and -0.03 for the u, g, r, i, and z filters, respectively \citep{chilingarian2011}. These corrections are far smaller than the observed asymmetry shown in this paper.



\section{ACKNOWLEDGMENT}

The research was supported in part by NSF grant IIS-1546079. I would like to thank Noah Brosch and Abraham Loeb for the fruitful discussions and kind assistance, as well as the associate editor Stephen Serjeant and the anonymous reviewer for their comments.
 
The Pan-STARRS1 Surveys (PS1) and the PS1 public science archive have been made possible through contributions by the Institute for Astronomy, the University of Hawaii, the Pan-STARRS Project Office, the Max-Planck Society and its participating institutes, the Max Planck Institute for Astronomy, Heidelberg and the Max Planck Institute for Extraterrestrial Physics, Garching, The Johns Hopkins University, Durham University, the University of Edinburgh, the Queen's University Belfast, the Harvard-Smithsonian Center for Astrophysics, the Las Cumbres Observatory Global Telescope Network Incorporated, the National Central University of Taiwan, the Space Telescope Science Institute, the National Aeronautics and Space Administration under Grant No. NNX08AR22G issued through the Planetary Science Division of the NASA Science Mission Directorate, the National Science Foundation Grant No. AST-1238877, the University of Maryland, Eotvos Lorand University (ELTE), the Los Alamos National Laboratory, and the Gordon and Betty Moore Foundation.

Funding for the SDSS and SDSS-II has been provided by the Alfred P. Sloan Foundation, the Participating Institutions, the National Science Foundation, the U.S. Department of Energy, the National Aeronautics and Space Administration, the Japanese Monbukagakusho, the Max Planck Society, and the Higher Education Funding Council for England. The SDSS Web Site is http://www.sdss.org/.

The SDSS is managed by the Astrophysical Research Consortium for the Participating Institutions. The Participating Institutions are the American Museum of Natural History, Astrophysical Institute Potsdam, University of Basel, University of Cambridge, Case Western Reserve University, University of Chicago, Drexel University, Fermilab, the Institute for Advanced Study, the Japan Participation Group, Johns Hopkins University, the Joint Institute for Nuclear Astrophysics, the Kavli Institute for Particle Astrophysics and Cosmology, the Korean Scientist Group, the Chinese Academy of Sciences (LAMOST), Los Alamos National Laboratory, the Max-Planck-Institute for Astronomy (MPIA), the Max-Planck-Institute for Astrophysics (MPA), New Mexico State University, Ohio State University, University of Pittsburgh, University of Portsmouth, Princeton University, the United States Naval Observatory, and the University of Washington.

\bibliographystyle{pasa-mnras}
\bibliography{galaxy_rotation_assym3}

\clearpage

\end{document}